# The influence of anisotropy on the evolution of interfacial morphologies in directional solidification: A phase-field study

Fengyi Yu (余枫怡)

CAS Key Laboratory of Mechanical Behavior and Design of Materials, Department of Modern Mechanics, University of Science and Technology of China, Hefei 230026, China

E-mail address: fengyi.yu.90@gmail.com

**ABSTRACT**

By adjusting the interface energy, curvature, and velocity, the anisotropy plays an important role in the interaction between interfacial processes and transport processes, determining the solidification structures. In this paper, through the quantitative phase-field model, the influence of anisotropy on the evolution of interfacial morphologies in directional solidification is investigated. To represent different interfacial anisotropies, the solidification processes with different preferred crystallographic orientations are performed. Then the effect of anisotropy on morphological evolution is discussed systematically. At the planar growth stage, the interfacial anisotropy makes no difference in the transport processes and morphological evolution. At the onset time of planar instability, the anisotropy determines the detailed evolution by adjusting the interface stiffness. At the planar-cellular-transition stage, with the influence of anisotropy, the interfacial curvature decreases from $\theta_0=0°$ to $\theta_0=40°$. Hence, the solute concentration ahead of the interface increases from $\theta_0=0°$ to $\theta_0=40°$, while the instantaneous velocity of the interface decreases from $\theta_0=0°$ to $\theta_0=40°$. At the quasi-steady-state stage, the anisotropy determines the growth direction and tip velocity of the primary dendrite, as well as the onset of sidebranches.

# I. INTRODUCTION

The mechanical properties of as-solidified parts are determined by the solidification structures. The precise control of structures requires a thorough understanding of solidification dynamics. Since different physical processes interact with each other at different scales in solidification [1-3], the investigation of solidification dynamics has been a long standing challenge. From the mesoscale viewpoint, solidification patterns result from the interaction between the interfacial processes and the transport processes of heat and mass [3-5]. By exchanging heat and mass with the environment, solidification patterns are formed out of equilibrium, which are dissipative structures created from irreversible processes [1,2]. The dissipative structures are represented by the evolution of interfacial morphologies, resulting in complex solidification structures, determining the defect formation and other properties of the components.

Due to the importance of interfacial morphologies, researchers developed various theoretical models to describe the evolution. For the planar interface instability, the descriptions go through the Constitutional Supercooling (CS) theory [6], Mullins-Sekerka (MS) analysis [7,8] to the Warren-Langer (WL) [9] model. The theoretical predictions of incubation time and average wavelength of planar instability consist with experimental observations [10,11]. For dendritic growth in undercooled melt, starting from the Ivantsov solution [12], the theories include the maximum velocity principle [13], marginal stability hypothesis [14], microscopic solvability condition theory [15,16], and the interfacial wave theory [17]. The theoretical models could identify the important parameters determining the evolution. However, the theoretical models involve many approximations and simplifications, resulting from the constraint that their solutions can only apply under the simple conditions. As a result, these models can hardly handle the complex morphologies of the interfaces and the relative interfacial effects. In the actual process, complex solidification structures result from the interaction between interfacial processes and transport processes. By adjusting the interface energy, relaxation time, curvature, and velocity, etc., the anisotropy plays an important role in the interaction [18]. Due to the constraint of the theoretical models, a systematic understanding of how anisotropy influences the evolution of the interfacial morphologies is still lacking.

As complementation to the theoretical models, the numerical models could be applied to the complex conditions, having the advantages of investigating the morphological evolution [19]. As one representative, the Phase-Field (PF) method combines the insights of thermodynamics and the dynamics of transport processes, having solid physical foundations [20-22]. Through the thin-interface asymptotic analysis, the dynamics of the interface in the PF model has been precisely controlled [22-24]. The governing equations provide a clear connection between anisotropy of the microscopic and macroscopic levels [18,25], both in equilibrium and non-equilibrium. By introducing the "Anti-Trapping Current" (ATC) term, the PF model could recover local equilibrium at the solid/liquid (S/L) interface and eliminate the spurious effects [26,27]. This PF model can predict the alloy solidification quantitatively, agreeing well with the experimental observations [10,28,29]. Moreover, as one of the interface capturing methods, the PF method uses an additional scalar field to implicitly represent the interface by one of its level sets. Hence it avoids the shape error caused by tracking interface in computation, having extremely high numerical accuracy. Hence, the PF method could capture the complex interfacial morphologies and the characteristic parameters at the interfaces [30,31]. It is suitable for representing the interaction between the interfacial processes and transport processes and investigating the effect of anisotropy on the morphological evolution.

In this paper, via the quantitative PF model, the influence of anisotropy on the evolution of interfacial morphologies in directional solidification is studied. To represent the different interfacial anisotropies, the solidification processes with different Preferred Crystallographic Orientations (PCOs) are carried out. Based on the simulations, the role of anisotropy in morphological evolution is discussed systematically.

## II. MODELS AND METHODOLOGY

### 1. Description of the model

The following is a brief introduction of the quantitative PF model for alloy solidification [26,27,32].

Firstly, a scalar variable $\phi(\mathbf{r}, t)$ is introduced to identify the state of phase, where $\phi=+1$ reflects the solid phase, $\phi=-1$ reflects the liquid phase, and intermediate values of $\phi$ reflects the S/L interface. Since $\phi$ varies smoothly across the interface, the sharp interface becomes diffuse and the phases turn into a continuous field, i.e., phase field $\phi(\mathbf{r}, t)$.

For the solute field, the composition $c(\mathbf{r}, t)$ is represented by the supersaturation field $U(\mathbf{r}, t)$, expressed by equation (1), where k is the solute partition coefficient, $c_\infty$ is the average solute concentration.

$$U = \frac{1}{1-k}\left(\frac{2kc/c_\infty}{1+k-(1-k)\cdot\phi}-1\right) \tag{1}$$

In directional solidification, the "frozen temperature approximation" is adopted, expressed by equation (2), where $T_0=T(z_0,0)$ is a reference temperature, $G(t)$ and $V_P(t)$ are the time-dependent thermal gradient and pulling speed. The approximation is on the basis of the assumptions: (1) The latent heat is ignored, i.e., the temperature field is undisturbed by interfacial evolution. It is essentially a statement concerning the relative magnitudes of the terms in the Stefan condition, $\rho_s L_f v^*_n \ll k_{s,l} \nabla T_{s,l} \cdot n$. (2) There is no flow in the liquid, consistent with the assumption that the densities of the solid and liquid are equal [33].

$$T(z,t) = T_0 + G(t)\left(z - z_0 - \int V_P(t)\,dt\right) \qquad (2)$$

Finally, the governing equations of the phase field and supersaturation field are given by [26,27]:

$$a_s^2(\hat{n})\left[1-(1-k)\frac{z-z_0-\int V_P(t)dt}{l_T}\right]\frac{\partial \phi}{\partial t} =$$
$$\nabla \cdot \left[a_s^2(\hat{n})\vec{\nabla}\phi\right] - \partial_x\left(a_s(\hat{n})\cdot a_s'(\hat{n})\cdot \partial_y \phi\right) + \partial_y\left(a_s(\hat{n})\cdot a_s'(\hat{n})\cdot \partial_x \phi\right) \qquad (3)$$
$$+ \phi(1-\phi^2) - \lambda(1-\phi^2)^2 \left[U + \frac{z-z_0-\int V_P(t)dt}{l_T}\right]$$

$$\left(\frac{1+k}{2} - \frac{1-k}{2}\phi\right)\frac{\partial U}{\partial t} = \nabla \cdot \left[D_L \cdot q(\phi) \cdot \vec{\nabla} U - \vec{j}_{at}\right] + \frac{1}{2}\left[1+(1-k)U\right]\frac{\partial \phi}{\partial t} \qquad (4)$$

where,

$$l_T = \frac{\Delta T_0}{G(t)} = \frac{|m|c_\infty(1-k)}{kG(t)} \qquad (5)$$

$$a_s(\hat{n}) \equiv a_s(\theta + \theta_0) = 1 + \varepsilon_4 \cos 4(\theta + \theta_0) \qquad (6)$$

$$\vec{j}_{at} = -\frac{1}{2\sqrt{2}}\left[1+(1-k)U\right]\frac{\partial \phi}{\partial t}\frac{\vec{\nabla}\phi}{|\vec{\nabla}\phi|} \qquad (7)$$

In the equations, $l_T$ is the thermal length, where m is the slope of the liquidus line in the phase diagram. $a_s(n)$ is the four-fold anisotropy function in a 2D system, where $\varepsilon_4$ is the anisotropy strength, $\theta$ the angle between the normal direction of the interface and the z-axis, $\theta_0$ is the intersection angle between the PCO of crystal and the z-axis. $D_L$ is the solute diffusion coefficient in the liquid. $q(\phi)=(1-\phi)/2$ is an interpolation function determining the varied diffusion coefficient across the domain. $\vec{j}_{at}$ is the ATC term, where $\partial\phi/\partial t$ reflects the rate of solidification, $\nabla\phi/|\nabla\phi|$ is the unit length along the normal direction of the S/L interface.

After ignoring the effect of kinetic undercooling, the calculation parameters in the governing equations could be linked to the physical qualities by the expressions: $W=d_0\lambda/a_1$ and $\tau_0=a_2\lambda W^2/D_L$, where W and $\tau_0$ represent the interface width and relaxation time, which are the length scale and time scale, respectively. In the expressions, $a_1=5\sqrt{2}/8$ and $a_2=47/75$, $\lambda$ is the coupling constant, $d_0=\Gamma/|m|(1-k)(c_\infty/k)$ is the chemical capillary length. $\Gamma=\gamma_{sl}T_f/(\rho_s L_f)$ is the Gibbs-Thomson coefficient, where $\gamma_{sl}$ is S/L interface energy, $T_f$ is the melting point of pure solvent and $L_f$ is the latent heat, respectively.

It should be noted, the governing equation (3) provides a clear connection between anisotropy of the microscopic and macroscopic levels [18,25]. Moreover, it unifies the anisotropy in equilibrium and non-equilibrium, revealing the interfacial anisotropy always plays an important role in solidification evolution.

## 2. Simulation procedure

The material parameters of Al-2.0wt.%Cu, regarded as a dilute binary alloy, are shown in Table 1 [28,34]. In the computation, the most important calculation parameter is the interface width W. The accuracy of the simulation increases with the decrease of W, while the computational cost increases dramatically with the decrease of W [23,24]. The thin interface limitation makes W just need to be one order of magnitude smaller than the characteristic length of the structure [27,35]. The characteristic length of alloy solidification is $L_C \sim \sqrt{d_0 * D_L/V_{tip}}$ [33], hence W was set to be 0.15μm. The periodic boundary conditions were loaded for the phase field and supersaturation field along the Thermal Gradient Direction (TGD). The time step size was chosen below the threshold of numerical instability for the diffusion equation, i.e., $\Delta t<(\Delta x)^2/(4D_L)$. Finally, this paper used fixed grid size $\Delta x=0.8W$ and time step size $\Delta t=0.013\tau_0$.

Table 1. The material parameters of Al-2.0wt.%Cu for the simulation [28,34]

| Symbol | Value | Unit |
| --- | --- | --- |
| Liquidus temperature, $T_L$ | 927.8 | K |
| Solidus temperature, $T_S$ | 896.8 | K |
| Diffusion coefficient in the liquid, $D_L$ | $3.0\times10^{-9}$ | m$^2$/s |
| Equilibrium partition coefficient, k | 0.14 | / |
| Alloy composition, $c_\infty$ | 2.0 | wt.% |
| Liquidus slope, m | -2.6 | K/wt.% |
| Gibbs-Thomson coefficient, $\Gamma$ | $2.4\times10^{-7}$ | K·m |
| Anisotropic strength of surface energy, $\varepsilon_4$ | 0.01 | / |

Moreover, to consider the infinitesimal perturbation of thermal noise on the S/L interface, a fluctuating current $J_U$ is introduced to the diffusion equation. By using the Euler explicit time scheme:

$$U^{t+\Delta t} = U^t + \Delta t\left(\partial_t U - \vec{\nabla}\cdot\vec{J}_U\right) \tag{8}$$

The components of $J_U$ are random variables obeying a Gaussian distribution, which has the maximum entropy relative to other probability distributions [36]:

$$\left\langle J_U^m(\vec{r},\vec{t}) J_U^n(\vec{r}\,',\vec{t}\,') \right\rangle = 2D_L q(\psi) F_U^0 \delta_{mn} \delta(\vec{r}-\vec{r}\,') \delta(t-t') \quad (9)$$

In equation (9), the constant noise magnitude $F_u^0$ means the value of $F_U$ for a reference planar interface at temperature $T_0$, defined as [37,38]:

$$F_U^0 = \frac{k v_0}{(1-k)^2 N_A c_\infty} \quad (10)$$

where $v_0$ is the molar volume of the solute atom, and $N_A$ is the Avogadro constant. Using the Clausius-Clapeyron relation:

$$\frac{|m|}{1-k} = \frac{k_B T_0^2}{\Delta h} \quad (11)$$

where $\Delta h$ is the latent heat per mole, and $k_B$ is the Boltzmann constant. The constant noise amplitude becomes:

$$F_U^0 = \frac{k}{|m| c_\infty (1-k)} \frac{k_B T_0^2}{L} \quad (12)$$

Finally, the program codes were written by C++. The explicit Finite Difference Method (FDM) was used when solving the governing equations, and the Message Passing Interface (MPI) parallelization was adopted for improving the computational efficiency.

### III. RESULTS AND DISCUSSION

In the computation, the dynamic solidification parameters are used. The thermal gradient G is constant 100K/mm, while the pulling speed $V_P$ increases from 0 to a fixed value 300μm/s, for which the increase time is 2.0s. The solidification processes with different intersection angles between the TGD and the PCO of crystal are simulated, where the angles are set to be 0°, 10°, 20°, 30°, and 40°. The computational domain is 2400×2400 grids, corresponding to 288.0μm×288.0μm in the real unit. It takes about 30 hours using 40 cores to finish one job.

**1. The planar growth**

The evolution of the characteristic parameters is shown in Fig. 1, including the solute concentration ahead of the interface $c_0$ and the instantaneous velocity of the interface $V_I$. Here $V_I$ is defined by $V_I=[z_0(t_2)-z_0(t_1)]/(t_2-t_1)$. Firstly, both $c_0$ and $V_I$ increase with time. At this stage, the $c_0$ curves of the different PCOs overlap with each other completely, so do the $V_I$ curves. The results indicate the interfacial anisotropy does not affect the solute diffusion at the planar growth stage, consistent with the literature [38]. As time goes on, the planar instability appears, represented by the sharp increase of the $V_I$ curves, shown in Fig. 1. The crossover times of the planar instability are the same between the simulations, i.e., the PCO of the crystal has little influence on the crossover time. According to the literature [39], the excess free energy $\Delta G$ and the corresponding interface energy $\gamma^0_{sl}$ are the critical parameters of the instability. When $\gamma^0_{sl}$ decreases to the critical level, with the influence of the anisotropy, the planar instability occurs. Here, in the simulations with the same solidification conditions and different PCOs, the characteristic parameters at the planar growth stage show the same tendencies, in Fig. 1, including the solute concentration $c_0$ and the instantaneous velocity $V_I$. Hence the changes of $\Delta G$ are the same between them, as well as the magnitude of $\gamma^0_{sl}$, determined by $\Delta G$. As a result, the crossover times are also the same between the simulations with different PCOs. That is, in the directional solidification, at the mesoscopic scale, the interfacial anisotropy only makes differences under the conditions the interfacial curvatures exist.

It should be noted, this result is different from that in the free growth process [18]. Since directional solidification is one kind of constrained growth process. To maintain the stable thermal gradient G and pulling speed $V_P$, the loaded parameters are dynamic but not static in the actual process. That is, the system we focus on is neither mass-conserved nor energy-conserved (open system, dissipative). In current study, the frozen temperature approximation is adopted. Hence the evolution of $c_0$ and $V_I$ are the same at the planar growth stage in the simulations with different PCOs. Fig. 2(a1)-(a5) shows the evolutional characteristics with different PCOs at the crossover time (t=1.44s), having the same distributions, including the diffusion length ahead of the interface, the pulling distance of the planar interface, the solute field and interfacial morphology. The results also demonstrate the previous conclusion that the anisotropy does not affect the planar growth.

In conclusion, at the planar growth stage, the anisotropy makes no difference in the transport processes and the evolution of interfacial morphologies.

## 2. The planar instability

Although the crossover times of the planar instability are the same in the simulations with different PCOs, the detailed characteristics differ with each other at this stage. The differences include the evolution of the solute concentration $c_0$ and instantaneous velocity $V_I$, shown in Fig. 1(a1)-(b1), as well as the amount of time for the instability, shown in Fig. 2(b1)-(b5). The distinctions result from the different PCOs of the crystal and the corresponding interfacial anisotropies. Specifically, when the interfacial curves exist, the anisotropy determines the solidification evolution by adjusting the surface energy and/or surface stiffness. The surface energy and surface stiffness are expressed in equations (13) and (14), where $\gamma^0_{sl}$ is the magnitude of the isotropic surface energy, determined by the excess free energy at the S/L interface.

$$\gamma_{sl} = \gamma^0_{sl}\left[1 + \varepsilon_4 \cos 4(\theta + \theta_0)\right] \tag{13}$$

$$\Psi_{sl} = \gamma_{sl} + \frac{d^2 \gamma_{sl}}{d\theta^2} = \gamma^0_{sl}\left[1 - 15\varepsilon_4 \cos 4(\theta + \theta_0)\right] \tag{14}$$

From the insight of thermodynamics, there are two common rules for the selection of growth direction: the maximum surface energy and the minimum surface stiffness. For the cubic crystal, the rule of maximum surface energy means the crystal will seek to minimize the total surface energy by creating large curvature in the <100> direction, while the rule of minimum surface stiffness means the crystal prefers to grow in the direction where the surface presents the smallest resistance to being deformed [33]. At the onset time of the planar instability, shown in Fig. 2(a), the magnitude of θ (the angle between the normal direction of the interface and the z-axis) is infinitesimal. Hence the surface stiffness in equation (14) can be simplified as $\Psi_{sl}=\gamma^0_{sl}[1-15\varepsilon_4\cos(4\theta_0)]$, indicating that the surface stiffness increases from $\theta_0=0°$ to $\theta_0=40°$. As a result, in Fig. 2(b1)-(b5), the cellular with $\theta_0=0°$ is the easiest to appear, taking the least time, while the cellular with $\theta_0=40°$ is the hardest to appear, taking the longest time. The phenomenon consists with the conclusion of Herring's relation [40], based on balance of free energies, illustrating the interfacial anisotropy determines the curvature undercooling by adjusting the interface stiffness.

In conclusion, if we define the incubation time as the onset time of the instability, the incubation times in the simulations with different PCOs are the same, in Fig. 2(a1)-(a5), consistent with [39]. If we define the incubation time as the time when the amplitude of the cellular becomes roughly comparable to its wave length, the incubation times increase from $\theta_0=0°$ to $\theta_0=40°$, in Fig. 2(b1)-(b5), consistent with the literatures [38,41]. The anisotropy determines the detailed evolution of the planar instability by adjusting the interface stiffness.

## 3. The planar-cellular-transition

As mentioned before, the interfacial anisotropy does not influence the evolution at the planar growth stage. By contrast, at the PCT stage, the appearing of the interfacial curvatures makes the anisotropy affect the evolution, including the solute concentration ahead of the interface $c_0$, the instantaneous velocity of the interface $V_I$, and the interfacial morphologies.

According to the sharp interface model (one-sided) of alloy solidification:

$$c_0 = \frac{c_\infty}{k} - \frac{\Gamma\kappa + Gz + \dot{T}t}{|m|} \qquad (15)$$

$$V_I = \left(-D_L \partial_n c \big|^+\right) / \left[(1-k)c_0\right] \qquad (16)$$

where $\kappa$ is the interfacial curvature, $\dot{T}$ is the cooling rate, and $\partial_n c|^+$ is the gradient of solute concentration at the liquid side of the interface. Note that $\partial_n c|^+ < 0$, to make the discussion intuitive, we set $|\partial_n c| = |\partial_n c|^+|$. Then,

$$V_I = D_L \cdot |\partial_n c| / \left[(1-k)c_0\right] \qquad (17)$$

Equation (15) illustrates $c_0$ is affected by the interfacial curvature $\kappa$. Equation (17) illustrates $V_I$ is dominated by $\partial_n c|^+$ and $c_0$, synergistically. Since the anisotropy directly influence the evolution of interfacial curvature $\kappa$, it will also impact the evolution of $c_0$, $V_I$, and the interfacial morphologies.

As shown in Fig. 3, at the PCT stage, on the one hand, according to the rule of maximum surface energy, to minimize the total surface energy, the crystal will seek to minimize the total surface energy by creating larger curvature in the <100> direction. On the other hand, due to the small distance between each cellular at this stage, the cellular growth is impressed significantly by the neighboring ones. Specifically, in Fig. 3(b1), the crystal with $\theta_0=0°$ could generate large curvature in the <100> direction. Because all the cellular grow along the direction of thermal gradient, the impression of the neighboring cellular has less impact on the tip curvature. By contrast, the crystal with $\theta_0=40°$ could hardly generate large curvature in the <100> direction directly. Because the distance of each cellular is quite small, the growth of the crystal along the <100> direction is restricted by the neighboring cellular. As a compromise, the crystal generates relatively small interfacial curvature, shown in Fig. 3(b5). That is, the curvature $\kappa$ decreases from $\theta_0=0°$ to $\theta_0=40°$, shown in Fig. 3 from (b1) to (b5). Since the solute concentration $c_0$ has a negative relation with the interfacial curvature $\kappa$, shown in equation (15). With the decrease of $\kappa$, from $\theta_0=0°$ to $\theta_0=40°$, $c_0$ increases from $\theta_0=0°$ to $\theta_0=40°$, shown by the $c_0$ curves after the crossover time (t=1.44s) in Fig. 1(b1). As for the instantaneous velocity $V_I$, equation (17) shows $V_I$ has a positive relation with $|\partial_n c|$ while it has a negative relation with $c_0$. The value of $|\partial_n c|$ decreases from $\theta_0=0°$ to $\theta_0=40°$, while $c_0$ increases from $\theta_0=0°$ to $\theta_0=40°$. Hence $V_I$ decreases from $\theta_0=0°$ to $\theta_0=40°$, shown by the $V_I$ curve in Fig. 1(a1).

In conclusion, at the PCT stage, with the influence of the anisotropy, the curvature $\kappa$ decreases from $\theta_0=0°$ to $\theta_0=40°$. As a result, $c_0$ increases from $\theta_0=0°$ to $\theta_0=40°$, while $V_I$ decreases from $\theta_0=0°$ to $\theta_0=40°$.

## 4. The onset of sidebranches

After the appearing of the cellular, in Fig. 3, a few sidebranches appear behind the tip of the primary dendrite, resulting from the decrease of the interface energy and stiffness induced by solute segregation [39]. Subsequently, with the solute concentration decreasing, fewer sidebranches grow out. Meanwhile, the onset of sidebranches corresponds to the large instantaneous velocity $V_I$, shown by the morphological evolution in Fig. 3(b1)-(b5). The complex dynamics at this stage results from the competitive effects of the interfacial curvature and velocity, depending strongly on the ratio $\tau/W^2$ in the PF model. The investigation will be carried out in the future.

As time goes further, solidification turns into the quasi-steady-state growth, shown by the stable curves (after t=3.5s) in Fig. 1 and the detailed evolution of interfacial morphologies and solute field in Fig. 4. According to [1,2], by exchanging heat and mass with the environment, solidification patterns are dissipative structures formed out of equilibrium. At this stage, the dissipative structures achieve a quasi-steady state after a period of self-organization. Hence the overall propagation velocities of the interface along the pulling direction are the same in the simulations, in Fig. 1(b). On the other hand, the anisotropy of interface energy determines the growth direction of the primary dendrite along the <100> direction, in Fig. 4. That is, although they have the same overall propagation velocities $V_I$, the dendrites with the different PCOs grow with the different tip velocities $V_{tip}$ along the <100> direction, where $V_{tip}$ is defined by $V_{tip}=[z_0(t_2)-z_0(t_1)]/(t_2-t_1)/\cos\varphi_0$. In the expression, $\varphi_0$ is the angle between the growth direction of the primary dendrite and the z-axis. Here we consider $\varphi_0$ equals to $\theta_0$. According to the expression of $V_{tip}$, the dendrites with the larger PCOs grow with the larger $V_{tip}$. In Fig. 4, the dendrites with larger PCOs have the larger $V_{tip}$ and tip curvature $\kappa$, which are more likely to grow out sidebranches. From the viewpoint of the whole domain, to keep the quasi-steady state of the dissipative structures, the system needs to exchange heat and mass with the environment. The larger $V_{tip}$ reflects the higher degree of non-equilibrium of the system, requiring more heat and mass exchange, bringing more non-equilibrium structures (sidebranches). Hence the dendrites with the larger PCOs are more likely to grow out sidebranches. From the viewpoint of the local domain, the onset of sidebranches can be regarded as one kind of interface instability, determined by the interface energy and its anisotropy. In Fig. 4, the sidebranches always appear behind the critical solute concentration, shown by the 4.5wt%Cu curves. The solute segregation decreases the interface energy. When the interface energy reduces to the critical level, with the influence of anisotropy, the instability occurs, resulting in the onset of sidebranches.

In conclusion, at the quasi-steady-state, the anisotropy determines the growth direction and tip velocity of the primary dendrite, as well as the onset of sidebranches, corresponding to large enough solute segregation (by decreasing the interface energy) and tip velocity (by increasing the degree of non-equilibrium).

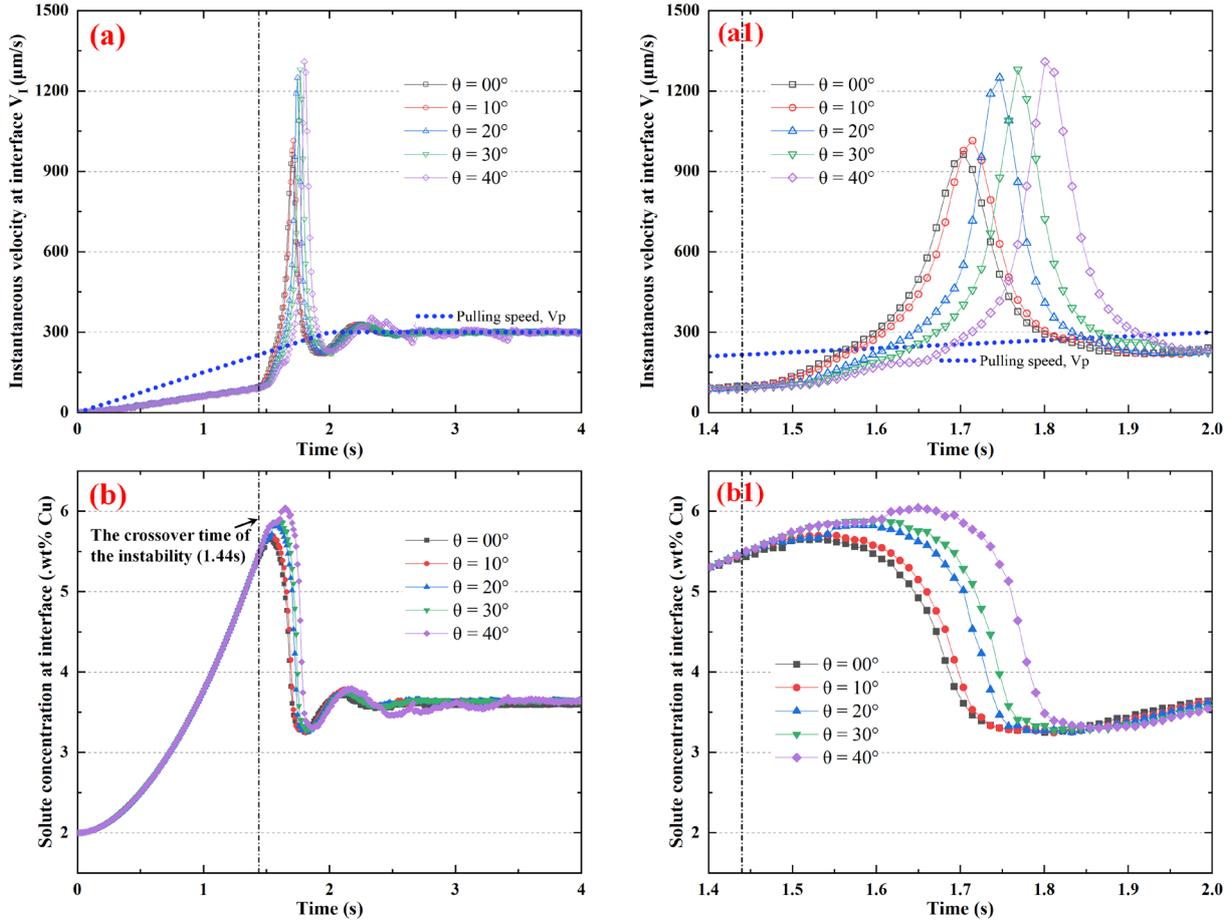

Fig. 1. The evolution of the characteristic parameters with different PCOs: (a) the instantaneous velocity of the interface $V_I$; (a1) is the enlarged version of (a); (b) the solute concentration ahead of the interface $c_0$; (b1) is the enlarged version of (b).

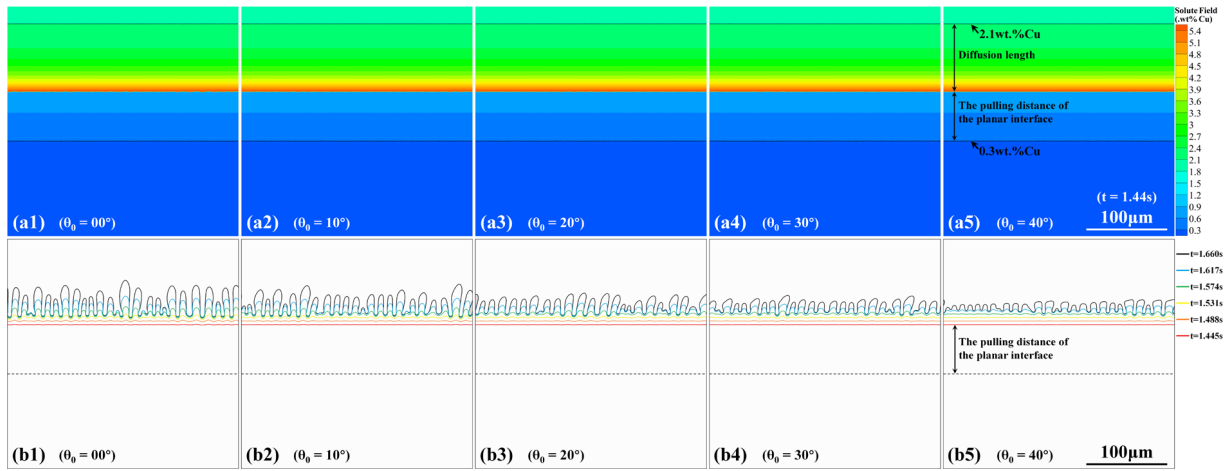

Fig. 2. (a) The interfacial morphology and solute field with different PCOs at the onset time of the planar instability (t=1.44s) and (b) the corresponding evolution of interfacial morphologies with different PCOs at the Planar-Cellular-Transition stage. (from the PF simulations)

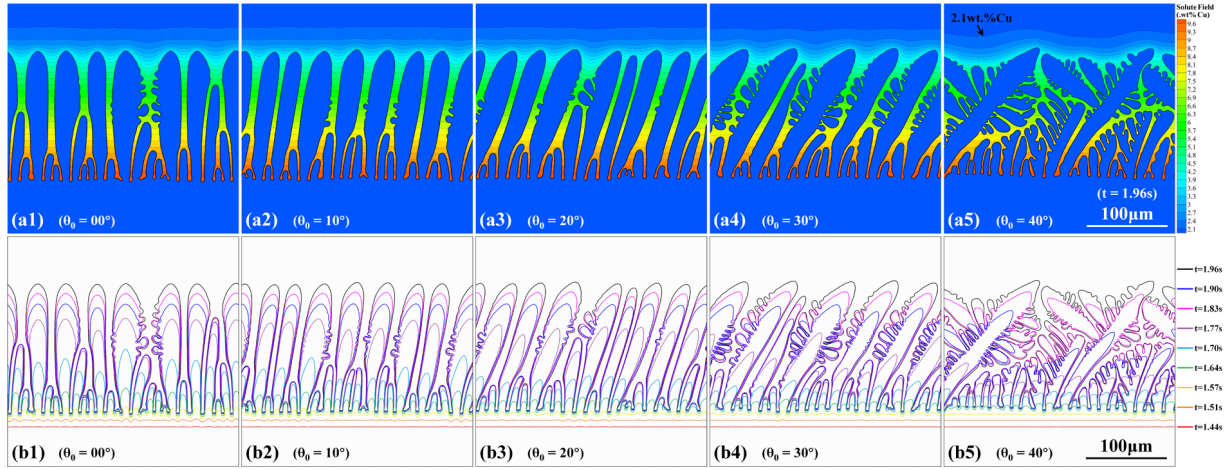

Fig. 3. (a) The interfacial morphology and solute field with different PCOs at the dendrite growth stage (t=1.96s) and (b) the corresponding evolution of interfacial morphologies with different PCOs at this stage. (from the PF simulations)

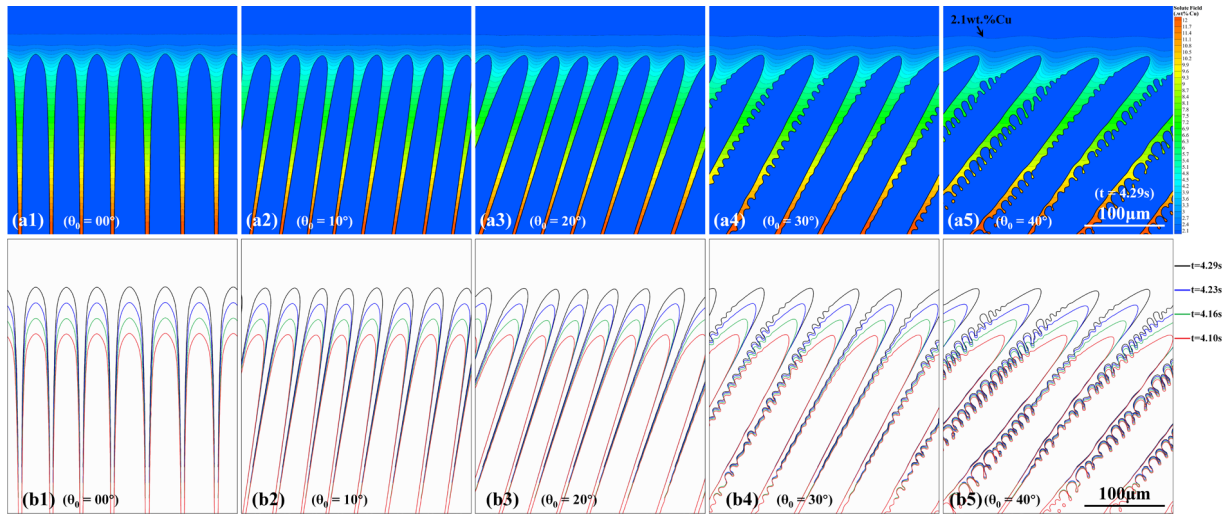

Fig. 4. (a) The interfacial morphology and solute field with different PCOs at the quasi-steady-state stage (t=4.29s) and (b) the corresponding evolution of interfacial morphologies with different PCOs at this stage. (from the PF simulations)

## IV. CONCLUSION

In this paper, using the quantitative PF model, the influence of anisotropy on the evolution of interfacial morphologies in directional solidification is investigated. To represent the different interfacial anisotropies, the solidification processes with different PCOs are performed. Then the influence of interfacial anisotropy on morphological evolution is discussed systematically, including the planar growth, the planar instability, the PCT stage, and the onset of sidebranches. The following conclusions could be drawn from the study:

(1) At the planar growth stage, the anisotropy makes no difference in the transport processes and the evolution of interfacial morphologies.

(2) If we define the incubation time as the onset time of the instability, the incubation times of different PCOs are the same. If we define the incubation time as the time when the amplitude of the cellular becomes roughly comparable to its wave length, the incubation times increase from $\theta_0=0°$ to $\theta_0=40°$. The anisotropy determines the detailed evolution of the planar instability by adjusting the interface stiffness.

(3) At the PCT stage, with the influence of the anisotropy, the curvature κ decreases from $\theta_0=0°$ to $\theta_0=40°$. As a result, $c_0$ increases from $\theta_0=0°$ to $\theta_0=40°$, while VI decreases from $\theta_0=0°$ to $\theta_0=40°$.

(4) At the quasi-steady-state, the anisotropy determines the growth direction and tip velocity of the primary dendrite, as well as the onset of sidebranches, corresponding to large enough solute segregation (by decreasing the interface energy) and tip velocity (by increasing the degree of non-equilibrium).

The dynamics of morphological evolution is determined by the competitive effects of the curvature (curvature-driven solute diffusion) and velocity, depending strongly on the ratio $\tau/W^2$ in the anisotropic PF model. It should be noted, the investigations here are based on the local equilibrium assumption. As the velocity increases further, this assumption is not suitable anymore. Moreover, due the increase of entropy at the interface, the solidification mode may turn into the facet growth. Then the role of anisotropy in the evolution should be considered through other ways.

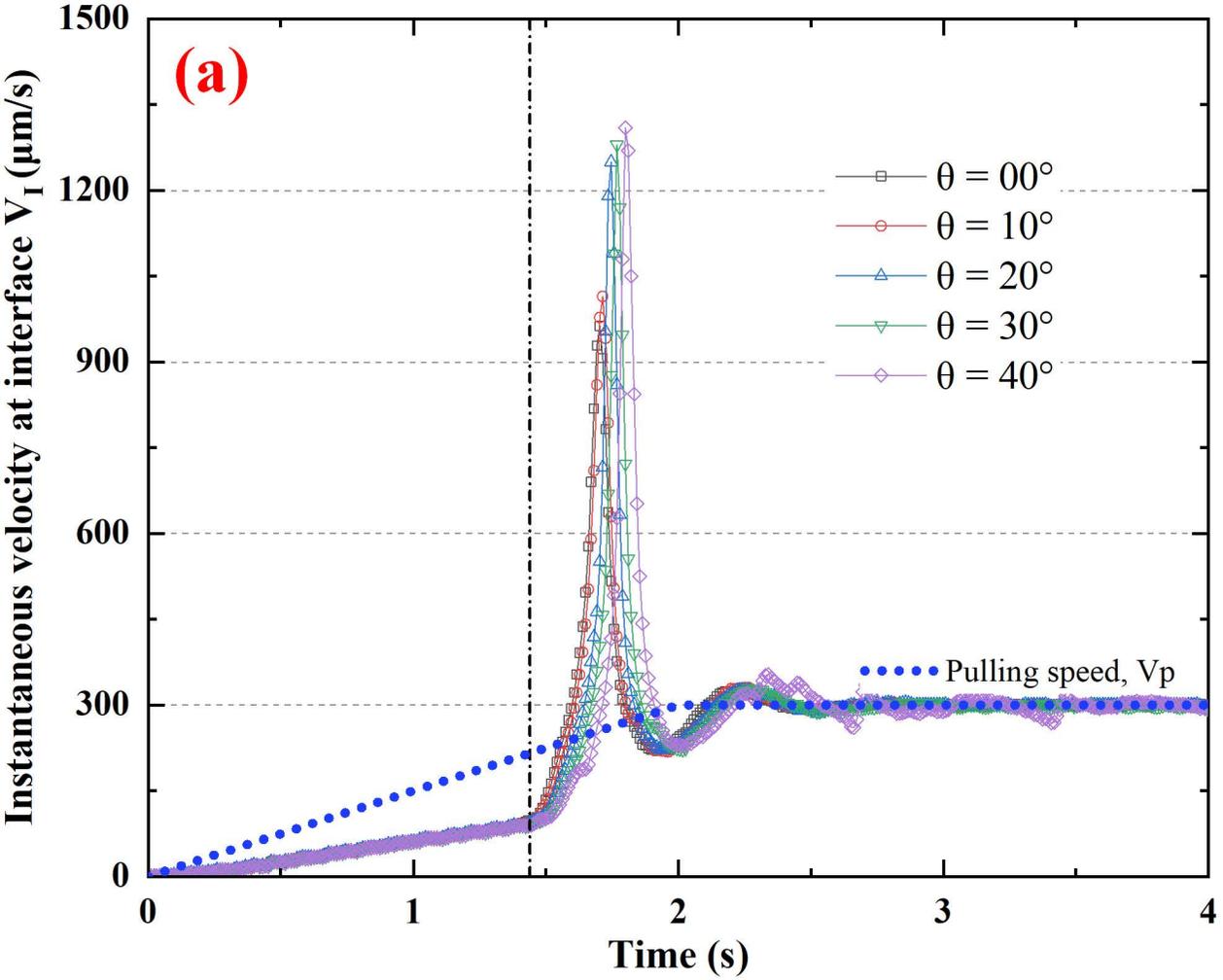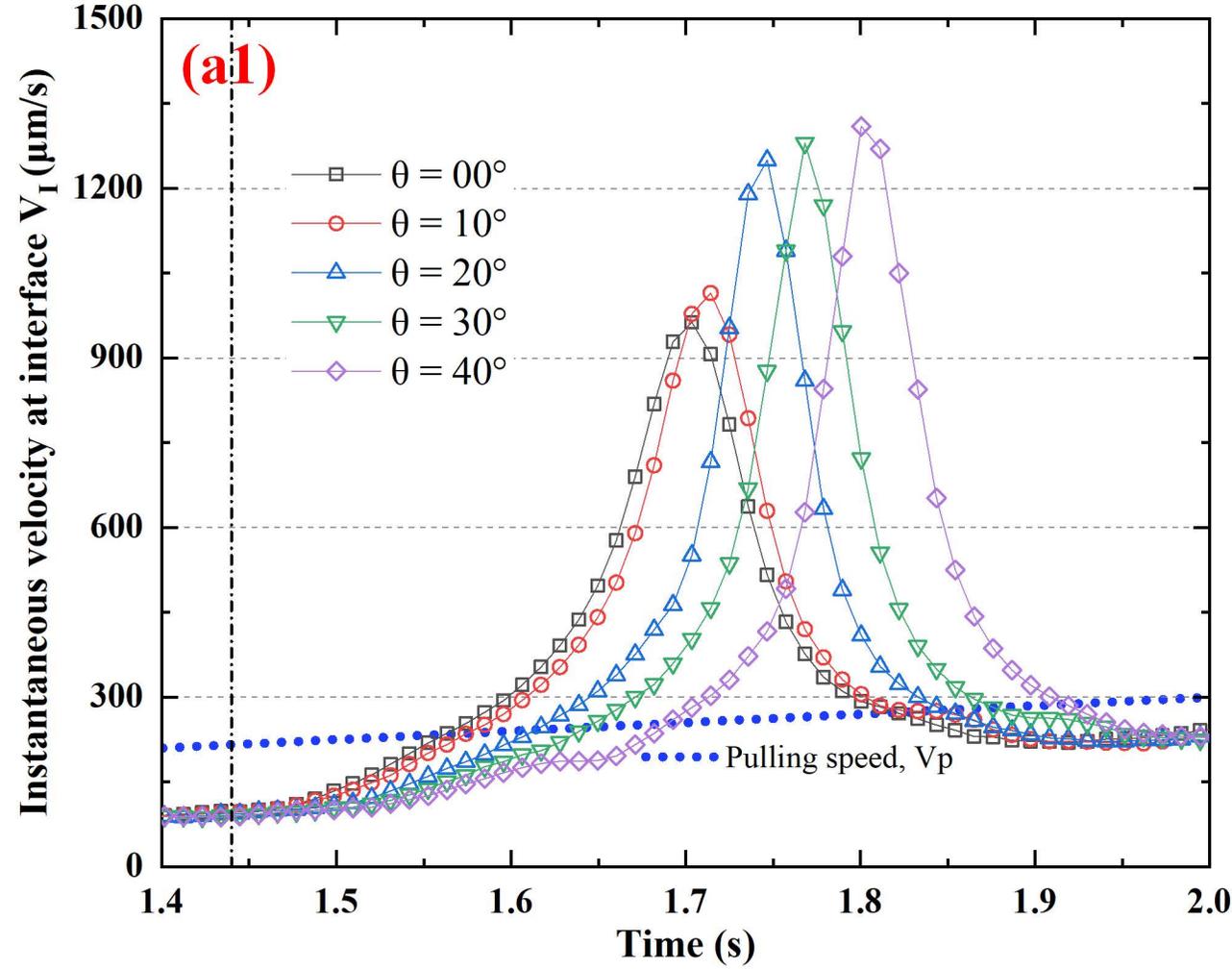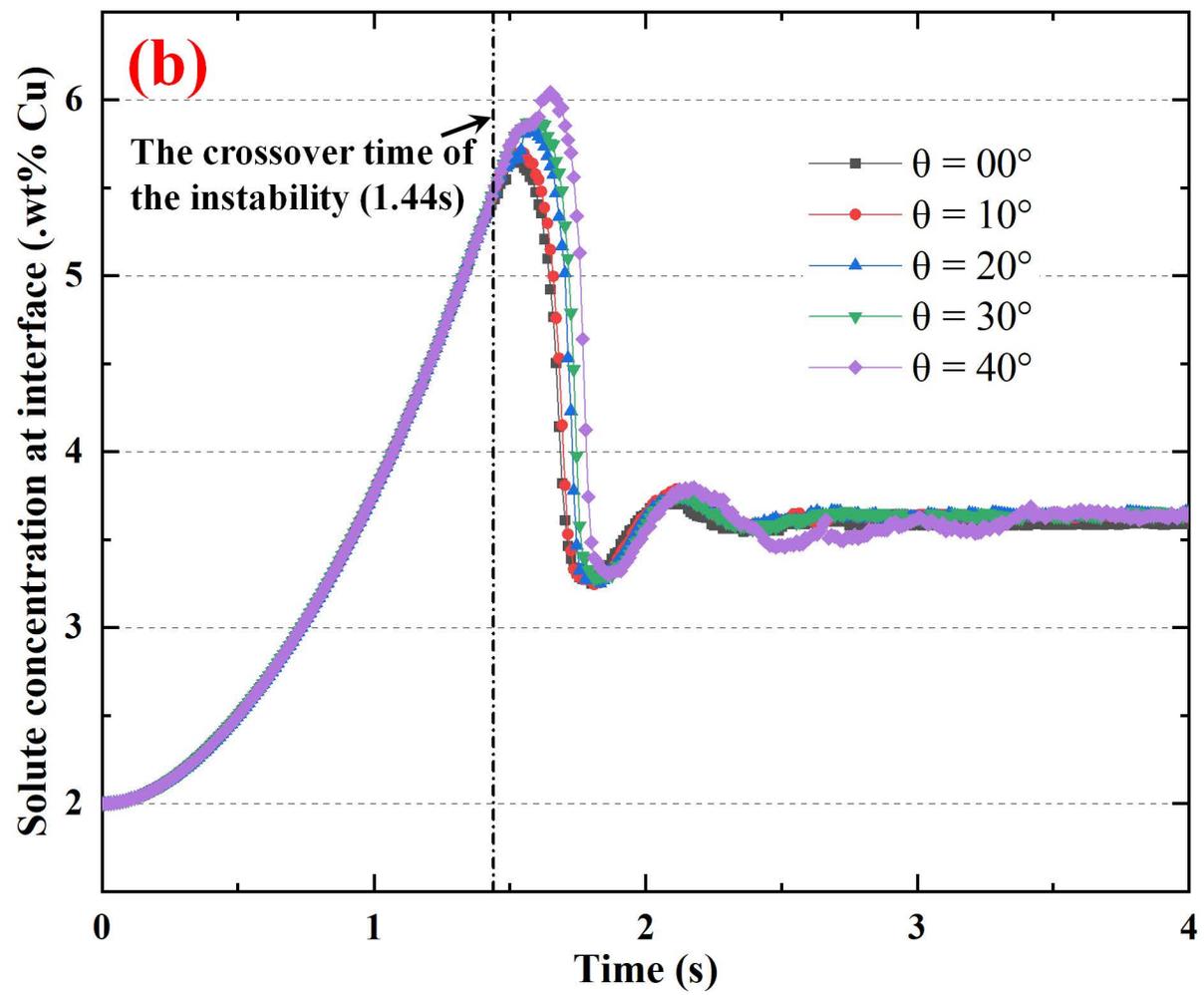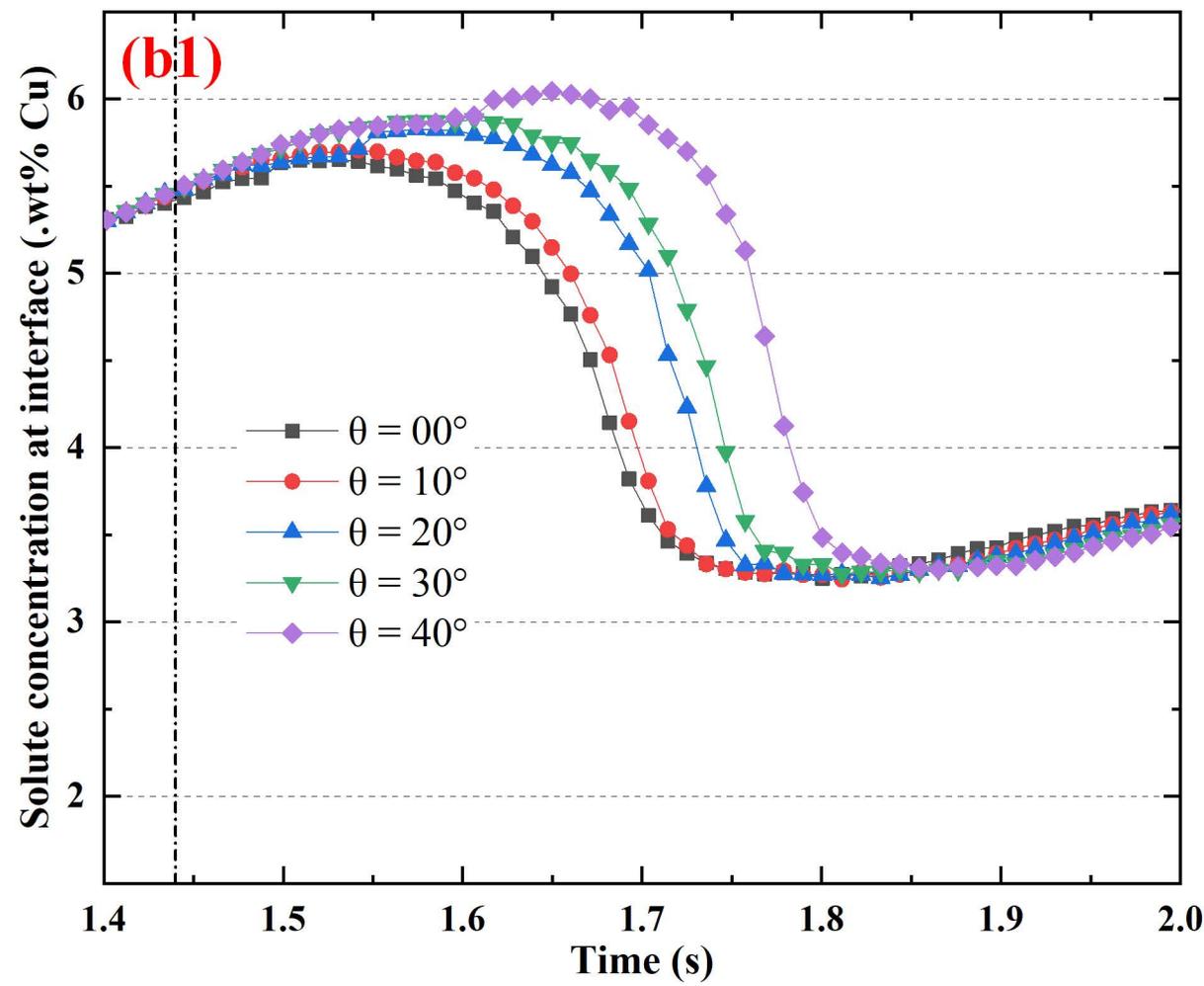

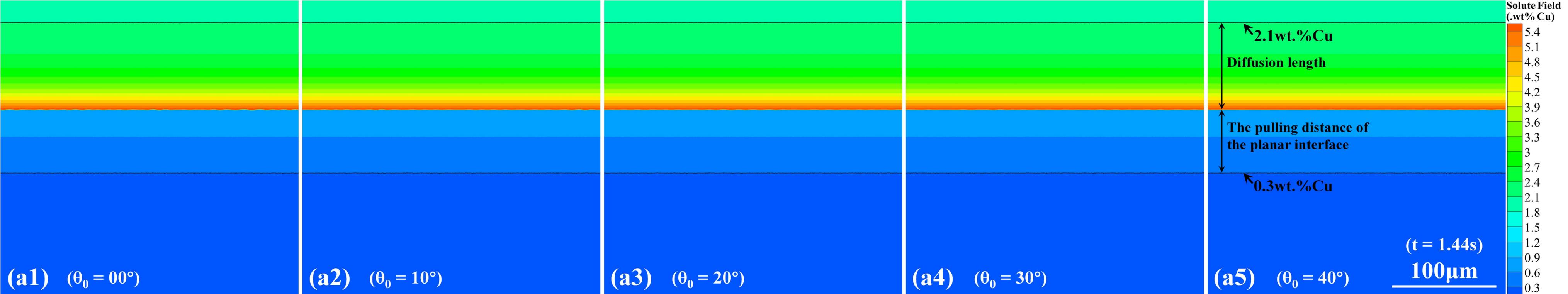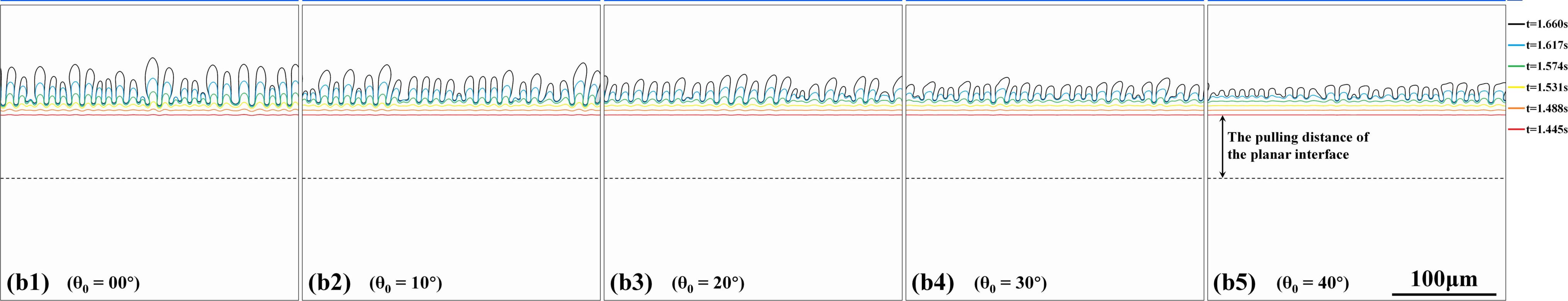

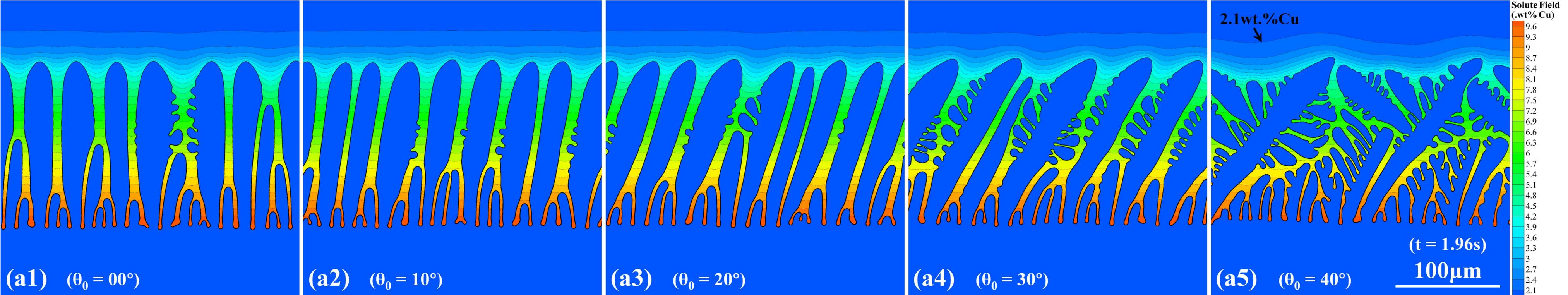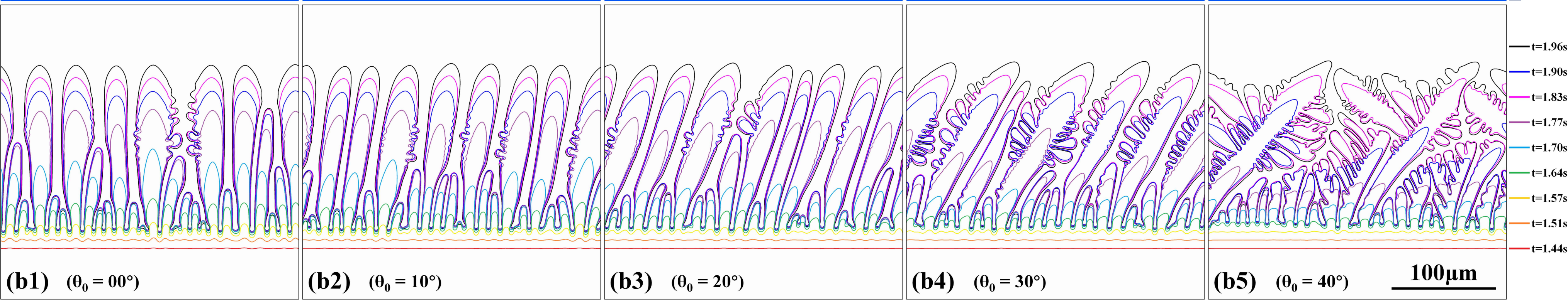

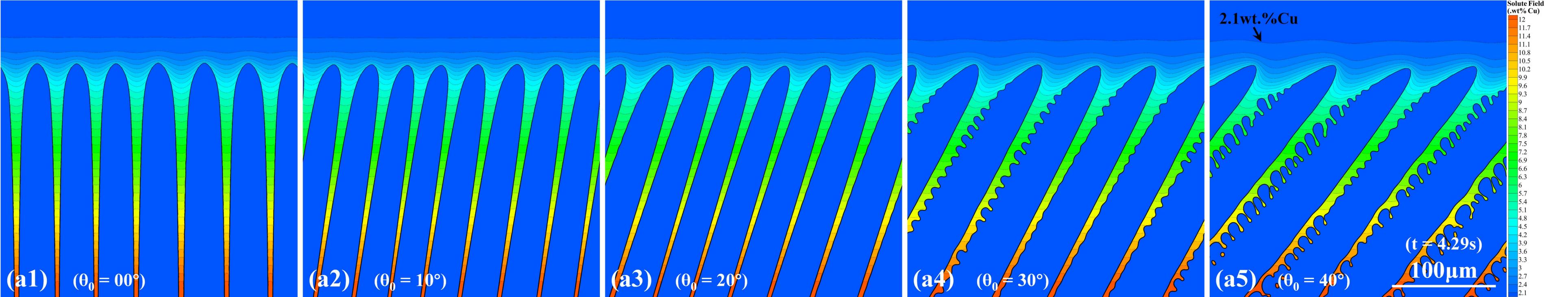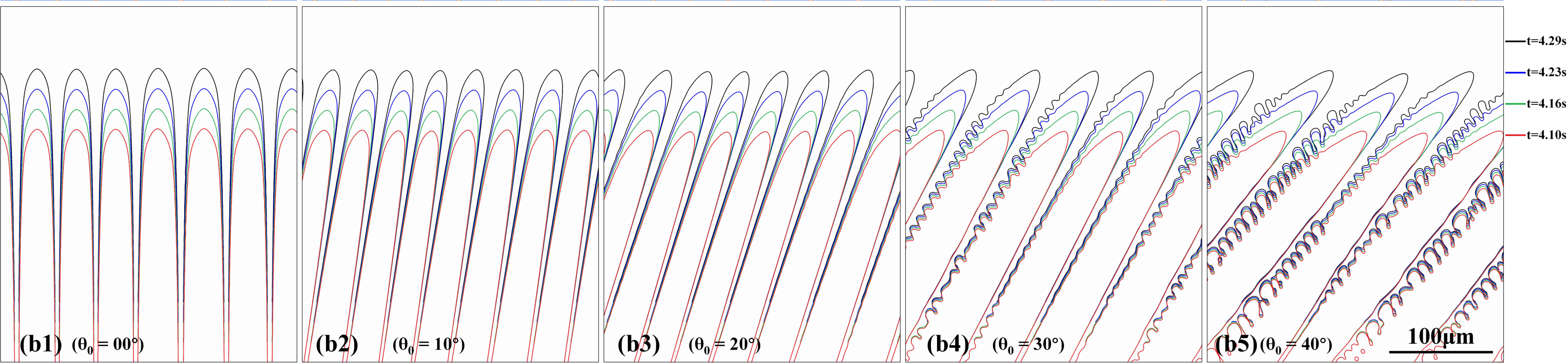